# How to Assess Trustworthy AI in Practice.

Roberto V. Zicari (1)(2)(3), Julia Amann (4), Frédérick Bruneault (5)(6), Megan Coffee (7), Boris Düdder (8), Alessio Gallucci (3), Thomas Krendl Gilbert (9), Thilo Hagendorff (10), Irmhild van Halem (3), Eleanore Hickman (11), Elisabeth Hildt (12)(1), Sune Holm (8), Georgios Kararigas (13), Pedro Kringen (3), Vince I. Madai (14)(15), Emilie Wiinblad Mathez (3), Jesmin Jahan Tithi (16)(17), Dennis Vetter (3)(18), Magnus Westerlund (1)(19), Renee Wurth (3). On behalf of the Z-Inspection® initiative (2022).

(1) Arcada University of Applied Sciences
(2) Seoul National University
(3) Z-Inspection® Initiative
(4) ETH Zurich
(5) Collège André-Laurendeau
(6) Université du Québec à Montréal
(7) New York University Grossman School of Medicine
(8) University of Copenhagen
(9) Cornell Tech
(10) University of Tuebingen
(11) University of Bristol
(12) Illinois Institute of Technology
(13) University of Iceland
(14) Charité Universitätsmedizin Berlin
(15) Birmingham City University
(16) Intel
(17) Stony Brook University
(18) Goethe University Frankfurt
(19) Kristiania University College

# Abstract

This report is a methodological reflection on Z-Inspection®.

Z-Inspection® is a holistic process used to evaluate the trustworthyness of AI-based technologies at different stages of the AI lifecycle. It focuses, in particular, on the identification and discussion of ethical issues and tensions through the elaboration of socio-technical scenarios. It uses the general European Union's High-Level Expert Group's (EU HLEG) guidelines for trustworthy AI.

This report illustrates for both AI researchers and AI practitioners how the EU HLEG guidelines for trustworthy AI can be applied in practice. We share the lessons learned from conducting a series of independent assessments to evaluate the trustworthiness of AI systems in healthcare. We also share key recommendations and practical suggestions on how to ensure a rigorous trustworthy AI assessment throughout the life-cycle of an AI system.

# Table of contents

# 1. Introduction

## 1.1 Trustworthy AI

Applications based on Machine Learning and/or Deep Learning carry specific (mostly unintentional) risks that are considered within AI ethics [1]. As a consequence, the quest for trustworthy AI has become a central issue for governance and technology impact assessment efforts, and has increased in the last four years, with focus on identifying both ethical and legal principles. Identifying such principles has become all the more important due to a tendency not just to use specific, single-purpose models, but rather large-scale "foundation models", which are trained on a broad range of data, fine-tuned on specific datasets, and eventually adapted to a wide range of downstream tasks [1].

As AI capabilities have grown exponentially, it has become increasingly difficult to determine whether their model outputs or system behaviors protect the rights and interests of an ever-wider group of stakeholders –let alone evaluate them as ethical or legal, or meeting goals of improving human welfare and freedom.

For example, what if decisions made using an AI-driven algorithm benefit some socially salient groups more than others? And what if we fail to identify and prevent these inequalities because we cannot explain how decisions were derived? Moreover, we also need to consider how the adoption of these new algorithms and the lack of knowledge and control over their inner workings may impact those in charge of making decisions.

Furthermore, there is a mismatch between the existence of high-level ethical guidelines and the practical implications for AI research and development. Four meta-studies have summarized nearly 100 high-level ethical guidelines for AI [2]–[8]. However, it has been noted that despite this inflation of guidelines for the ethical conduct of AI research, there is no shortage of reports of unethical applications of AI [4]. The main reason for this is that the current frameworks are very abstract and have limited practical application for researchers and developers of algorithms and AI systems [4]. One recent study found that 75% of the main ethical guidelines only contain general principles with very little detail, and over 80% offer no or very little practical insights [9].

This contradiction between having both an abundance of high-level guidance and unethical applications of AI is partly reflected in the need for systematic auditing methods aimed at assessing or increasing the trustworthiness of AI applications. These methods typically cover (preventive) measures to render AI applications fair, to ensure compliance with privacy and safety requirements, to increase technical explainability and further dimensions of transparency, to anticipate potential social ramifications, and many more. Furthermore, the credibility or trustworthiness of AI applications must either result from firsthand experiences, or be reputed by third-party assessments, the latter being the most promising and reliable approach [10].

This is where Z-Inspection® [11] comes into play, either as a co-design, self-assessment, or auditing method. Its ultimate goal is to foster high levels of trustworthiness of AI systems, entailing them to be fair, safe, transparent, as well as socially acceptable. Z-Inspection® is a holistic process based on the method of evaluating new technologies, where ethical issues need to be discussed through the elaboration of socio-technical scenarios [11]. In particular, Z-Inspection® can be used to perform independent assessments and/or self-assessments together with the stakeholders owning the use case.

This report is a methodological reflection on Z-Inspection®; it illustrates for both AI researchers and AI practitioners how the EU HLEG guidelines for trustworthy AI can be applied in practice.

The key question is: How do we know whether and for whom AI is "good"? To try to answer this question, we have conducted a number of assessments for trustworthy AI , primarily in the field of medicine and healthcare using the Z-Inspection® process (see Section 2).

In this paper, we also share key recommendations on how to ensure a rigorous trustworthy AI assessment during the full life-cycle of an AI system.

## 1.2 Using the EU Framework for Trustworthy Artificial Intelligence

For the context of our work we define ethics in line with the essence of modern democracy i.e. "respect for others, expressed through support for fundamental human rights" [12]. We take into consideration that "*trust*" in the development, deployment and use of AI systems concerns not only the technology's inherent properties, but also the qualities of the socio-technical systems involving AI applications" [13].

Specifically, we consider the ethics *guidelines* for *trustworthy* artificial intelligence defined by the EU High-Level Expert Group on AI, which defined *trustworthy AI* [13] as:

(1) **lawful** - respecting all applicable laws and regulations
(2) **ethical** - respecting ethical principles and values
(3) **robust** - both from a technical perspective and taking into account its social environment

And we use the four ethical principles, rooted in fundamental rights defined in [13], acknowledging that tensions may arise between them:

(1) **Respect for human autonomy**
(2) **Prevention of harm**
(3) **Fairness**
(4) **Explicability**

Furthermore, we also consider the seven requirements of Trustworthy AI defined in [13]. Each requirement has a number of sub-requirements as indicated in Table 1.

Table 1. Requirements and sub-requirements Trustworthy AI. Reproduced from [13].

| Requirements | Sub-Requirements |
| --- | --- |
| 1 Human agency and oversight | *Including fundamental rights, human agency and human oversight* |
| 2 Technical robustness and safety | *Including resilience to attack and security, fall back plan and general safety, accuracy, reliability and reproducibility* |
| 3 Privacy and data governance | *Including respect for privacy, quality and integrity of data, and access to data* |
| 4 Transparency | *Including traceability, explainability and communication* |
| 5 Diversity, non-discrimination and fairness | *Including the avoidance of unfair bias, accessibility and universal design, and stakeholder participation* |
| 6 Societal and environmental wellbeing | *Including sustainability and environmental friendliness, social impact, society and democracy* |
| 7 Accountability | *Including auditability, minimization and reporting of negative impact, trade-offs and redress.* |

While we consider the seven requirements comprehensive, we believe additional ones can still bring value. Two of such additional requirements proposed by the Z-Inspection® initiative are "*Assessing if the ecosystems respect values of Western Modern democracy*" and "*Avoiding

*concentration of power"* [13]. We will discuss in Sec. 2.1.4 how these two additional requirements play a key role in the evaluation of the boundaries for the assessment.

## 1.3 Challenges and Limitations of the EU Framework for Trustworthy AI

The AI HLEG trustworthy AI *guidelines* were formulated as non-legal and non-binding guidance to direct the development of AI towards the consideration of a wide range of ethical principles in a bid to balance innovation with safety [14]. Given the broad scope of AI systems and that the meaning of AI itself is still a matter of debate, the seven requirements laid out in the guidelines have not been anchored to a specific context [15]. These guidelines have formed the foundation upon which a proposed Artificial Intelligence Act (AIA) has been built [16]. The AIA is still in early European Commission procedural stages but, as drafted, it categorizes AI systems (defined broadly) into those that are: prohibited, of unacceptable, high risk, limited risk and low risk [17]. Those AI systems categorized as high risk will need to fulfill a number of requirements before being used or placed on the market. Many of those requirements correspond with aspects of the guidelines, for example transparency and human oversight. Assessment as to whether those requirements are fulfilled by specific AI systems will be carried out by any third party bodies who are already responsible under pre-existing product safety legislation (i.e. such as there is in relation to any medical devices being brought to market in the EU, under the MDR). Currently the AIA proposes that AI that is not subject to pre-existing legislation will need to have its conformity with the requirements self-assessed by those responsible for it.

While the AIA is being developed, AI systems remain covered by the framework of the guidelines with which there are numerous issues. For instance, the guidelines do not reflect whether it is reasonable to use AI in the first place to address or solve a particular (social) problem [18], hence falling prey to the "framing trap" [19]. This means it remains possible for systems optimized in terms of fairness, explainability, safety, etc., to cause harm that is not ascertainable by these principles [20]. Even though the EU framework makes explicit reference to the social robustness of AI systems (something most other guidelines ignore), we believe that this element should play a more important role in assessing AI systems. The guidelines also do not discuss the fact that trustworthy AI is less about AI models themselves than about the socio-technical contexts in which AI systems operate and may transform, including data collection and digital infrastructure, market power, industry competitiveness, and the like. Hence, there is the risk that many recommendations will be non-actionable due to conflicting monetary interests and incentives. We wait to see how the legislation will deal with these, and other issues but, no matter how much legislation there will be, there will still be many aspects of AI application that will be governed by ethics. High level principles do not guarantee practically applied ethics.

In connection with the HLEG guidelines, the EU framework offers a static checklist and a web tool (The Assessment List for Trustworthy Artificial Intelligence (ALTAI)) [21], [22] designed to enable self-assessment of the trustworthiness of AI systems. The ALTAI assessment provides no validation of claims regarding trustworthiness and does not take into account changes of AI technology over time.

## 2. How to assess Trustworthy AI in *practice*?

Despite the challenges and limitations of the AI HLEG, trustworthy AI guidelines comprise a cohesive framework for conducting a self-assessment of AI systems. These underpinning principles are very useful in organizing a systematic assessment of AI systems, especially compared to other guidelines available, which do not detail specific requirements.

That is why we use the main principles and the 7 requirements for a trustworthy AI in the Z-Inspection® process. In our research work, we looked at how to apply these principles in practice. However, the EU guidelines require implementation by those with broad and deep knowledge of the subject area environment. The potential concerns regarding effectiveness, unintended impacts, and inequities require more than a one-size-fits all evaluation, but requires independent interdisciplinary experts with specific knowledge bases. This has been the strength of the Z-Inspection® process and something guidelines alone cannot supply.

In this section, and in the rest of the report, we reflect on the methodology of Z-Inspection® in terms of both its use of the EU HLEG AI guidelines and its application to real-world use cases. We share the lessons learned in conducting a number of self-assessments to evaluate the trustworthiness of AI systems in healthcare.

## 2.1 Z-Inspection® Process in a Nutshell

We created an *orchestration process,* called Z-Inspection®, to help teams of skilled experts to assess the *ethical, technical, domain-specific* and *legal* implications of the use of an AI-product/service within given *contexts* [11].

The process is composed of three phases (see Figure 1): set up, assess and resolve.

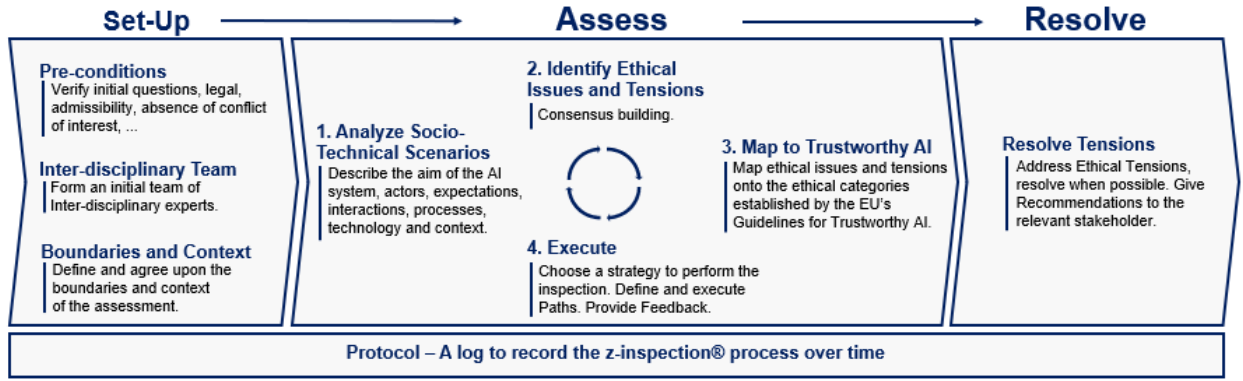

Figure 1. Z-Inspection® Process flow chart describing the main steps of the Set-Up, Assess and Resolve phase. Additionally, a log is kept in which the process and events of the assessment are tracked. Adapted from [11].

The *Set Up* phase consists of the validation of a number of pre-conditions before the assessment starts, the set up of an interdisciplinary team of experts working together with the key stakeholders owning the specific AI use case, and finally the definition of the boundaries and context where the assessment takes place.

The *Assess phase* is an iterative process that includes the creation and analysis of socio-technical scenarios, the identification of ethical issues and tension, the validations of claims by providing evidence (if any) and the mapping to the EU trustworthy AI framework using a mapping from "open to closed vocabulary" as a consensus based approach .

The *Resolve phase* addresses the ethical tensions identified during the Assess phase. Where possible trade-off solutions are proposed, possible risks and remedies are identified, and recommendations are made to the key stakeholders.

A detailed description of the three phases can be found in Zicari et al. [11].

The Z-Inspection® Process can be applied to the Entire AI Life Cycle: i) Design, ii) Development, iii) Deployment , iv) Monitoring, and v) Decommissioning.

In the design phase, the process can provide insight on how to design a trustworthy system.

During the development phase, the process can be used to specify test cases, e.g., to verify the absence of certain biases, especially when requirement changes or refinement happens in the original requirements in the fourdesign.

During the deployment, the process can be integrated into the acceptance test if trustworthiness is a specified user requirement. Since AI systems evolve over time due to updated models, algorithms, data, or environments, the trustworthiness needs to be assessed as a continuous monitoring process. The decommissioning of systems and replacement by other systems is a critical activity in systems due to the required compatibility for other systems using the functionality of the soon-to-be-replaced AI system. Here, the protocols and logs (recording document) of the process over the full lifecycle of the old trustworthy AI system can facilitate the lifecycle of the new product.

In addition to the application of the Z-Inspection® at the design, deployment and evaluation phase, it should be considered whether an AI system is the most appropriate solution to use in a particular case. Development of regulatory guidance pertaining to whether it is reasonable to use AI is currently lacking and would contribute to mitigating risks to democratic values and human rights and ensuring that AI is deployed where it has potential to contribute for the benefit of those affected. Regulatory guidance on when to use AI would complement assessments of the AI once in use, such as the Z-Inspection®.

### 2.1.1 Pre-Conditions

When assessing trustworthy AI, as part of the Set Up phase, we suggest verifying some pre-conditions, namely:

- Who requested the inspection?
- Why carry out an inspection?
- For whom is the inspection relevant?
- Is it recommended or required (mandatory inspection)?
- What are the sufficient vs. necessary conditions that need to be analyzed?
- How are the inspection results expected to be used?
- Will the results be shared to the public domain or kept private?
- Are there any conflicts of interest (for both the stakeholders owners of the AI and for the team of independent experts)
- Clarify *what is* and *how to handle* the *IP* of the AI and of the part of the entity/company to be examined.
- Identify possible restrictions to the Inspection process, in this case, assess the consequences (if any).
- Define if and when *Code Reviews* are needed/possible. For example, check the following preconditions [23]:
    - There are no risks to the security of the system
    - Privacy of underlying data is ensured
    - No undermining of intellectual property
- Define the implications if any of the above conditions are not satisfied. For example:
    - Which stakeholders (if any) have been left out of scope? For what reason(s)?
    - Between participating experts , how will conflicts of interest be addressed?

- Will the inspection be revisited at a later date? Will the participants change?

### 2.1.2 Best Practices

We have used Z-Inspection® to assess trustworthy AI for four healthcare use cases. We have considered AI models in different stages of their development and different requirements giving the opportunity to further improve and develop the initial Z-Inspection® process with real world complex examples. Our previous collaborations include evaluating: (1) a deployed ML pipeline estimating the risk of cardiovascular disease [11]; (2) a deployed ML model that supported a tool for detection of cardiac arrest in emergency calls [15], [24]; (3) a co-design of a Deep Learning-based tool to help dermatologists detect malignancy in skin lesions [25], [26]; and (4) a Deep Learning-based system for screening pulmonary findings in COVID-19 patients (in revision) [52]. We will elaborate more for all use cases and the lessons learned in the following sections. For all use cases, the assessment work we performed was conducted purely for research purposes and did not involve any compensation or personal benefits.

### i) **AI For Predicting Cardiovascular Risk**

*The Problem Domain:* Cardiovascular diseases (CVD) are the leading cause of morbidity and mortality worldwide, accounting for an estimated 17.9 million lives annually [27]. Several ML approaches have been developed for cardiovascular disease diagnosis and prediction [28]. The potential of AI in cardiovascular medicine is high; however, ignorance of the ethical challenges may overshadow its potential clinical impact.

*Use Case:* We have used and tested Z-Inspection® by evaluating a ML-based screening tool to assist medical doctors in the detection of coronary ischemia in a non-invasive manner, using ML to analyze sensor data (i.e. electrical signals of the heart). The company that developed this used a traditional ML pipeline approach, which transforms the raw data into features that better represent the predictive task. The features are interpretable and the role of ML is to map the representation to output [11].

*Status:* The assessment was conducted with the AI system already deployed and in use in several countries in Europe and in other parts of the world. The single-centered, Euro-centric development of this AI tool made it important to consider the ethics of applying such a tool with populations in other geographical areas, which may have important differences from the original cohort used for training. This important aspect was not taken into account by the tool developers when building or testing it [11].

### ii) **Machine Learning as a Supportive Tool to Recognize Cardiac Arrest in Emergency Calls**

*The problem:* Health-related emergency calls (112) are part of the Emergency Medical Dispatch Center (EMS) of the City of Copenhagen, triaged by medical dispatchers (i.e., medically-trained dispatchers, who answer the call, e.g., nurses and paramedics) and medical control by a physician on-site. In recent years, the Emergency Medical Dispatch Center of the City of Copenhagen has been unable to identify approximately 25% of cases of out-of-hospital cardiac arrest (OHCA) [24], [29]. These cases have only been recognized once the paramedics/ambulance arrived at the scene. Therefore, the Emergency Medical Dispatch Center of the City of Copenhagen loses the opportunity to provide the caller with instructions for cardiopulmonary resuscitation (CPR), and hence, impairs survival rates.

*Use Case:* A team led by Stig Nikolaj Blomberg (Emergency Medical Services Copenhagen, and Department of Clinical Medicine, University of Copenhagen, Denmark) worked together with a start-up company and examined whether a ML framework could be used to recognize out-of-hospital cardiac arrest (OHCA) cases by listening to the calls made to the Emergency Medical Dispatch Center of the City of Copenhagen. The company designed and implemented the AI system, and trained and tested it by using the archive of audio files of emergency calls provided by Emergency Medical Services Copenhagen in the year 2014. The prime aim of this AI system is to assist medical dispatchers when answering 112 emergency calls to help them to early detect OHCA during the calls, and, therefore, possibly saving lives.

*Status:* The AI system was put into production during Fall 2020.

The research questions we addressed were: Is the AI system trustworthy? Is the use of this AI system trustworthy? For that, we conducted a self-assessment conducted jointly by a team of independent experts together with the prime stakeholders of this use case. The main motivation of this work was to identify possible risks and pitfalls of using the AI system assessed here, and to provide recommendations to the key stakeholders. The work was done in cooperation with Emergency Medical Services Copenhagen, and the Department of Clinical Medicine, University of Copenhagen, Denmark [15], [30].

The assessment was conducted with the AI system already deployed and in use in the city of Copenhagen. One of the main findings of the assessment was that the AI system was designed, trained and tested using rather homogenous data sets, resulting in the risk that when people speak Danish with an accent, the system may not be accurate. This risk was not considered at the time the decision for deployment was made.

Among other recommendations, we suggested the following :

Recommendation 3: Involve stakeholders. The group of (potential future) patients and (potential future) callers could be interested in knowing how the system functions and is developed. User /stakeholder involvement is likely to be helpful in the process of redesigning the AI system.

Recommendation 4: It is important to learn how the protocol (what questions, how many, etc.) does or does not influence the accuracy of the ML output. Further research work should be performed to answer this question. The goal should be to responsibly integrate the classifier into the context of the dispatcher calls rather than just have it passively observe the call and make "trustworthy" recommendations. This requires reimagining the context of the calls themselves (with new protocols, questions, etc.).

iii) **Deep Learning based Skin Lesion Classifier.**

*The Problem domain:* Skin cancer is primarily diagnosed visually by dermatologists, beginning with an initial clinical screening and followed potentially by dermoscopic analysis, a biopsy and histopathological examination [53]

*Use Case:* A team of researchers at the German Center for Artificial Intelligence (DFKI) developed an AI framework that not only classified skin lesions, but included descriptive outputs to facilitate learning among dermatologists in training.

Automated classification of skin lesions using images is a challenging task owing to the fine-grained variability in the appearance of skin lesions. [53]

The initial design aim of the AI system was to act as an add-on component to support dermatologists in clinical practice. Given an existing AI system trained for skin melanoma detection, the add-on component goal was to explain the system's decisions in terms that dermatologists can understand. With these explanations, the AI system could support the clinician's decision-making process by providing a qualified second opinion on relevant features in a dermatoscopic image and, therefore, potentially improve diagnostic performance.

Deep Learning classifiers for skin lesions are often difficult to interpret and understand. Several methods have been proposed to make them more transparent, for example in the field of Explainable AI. Exampling its decision might facilitate AI adoption by the general public or facilitate in their daily work or training.

*Status:* We used Z-Inspection® as an ethically aligned **co-design** methodology to ensure trustworthiness in the early design phase of the above system. The system explains decisions made by Deep Learning networks analyzing images of skin lesions. The co-design of trustworthy AI developed used a holistic (comprehensive) approach rather than a static ethical checklist and required an interdisciplinary team of experts working with the AI designers and their managers. Ethical, legal, and technical issues potentially arising from the future use of the AI system were investigated. One of the key findings in using the assessment in early design is that by including different viewpoints from domain experts it has an impact on the overall design of the AI system. Our results can also serve as guidance for other similar early-phase AI tool developments [25].

## iv) Deep Learning for predicting a multi-regional score conveying the degree of lung compromise

*The problem domain:* Worldwide, the saturation of healthcare facilities, due to SARS-CoV-2 infections and the significant rate of respiratory complications, has been a critical aspect of the COVID-19 pandemic.

*Use case:* The team of Alberto Signoroni, Davide Farina and colleagues at the University of Brescia and the Brescia Public Hospital (ASST Spedali Civili) implemented an end-to-end deep learning architecture, designed for predicting, on Chest X-rays images (CXR), a multi-regional score conveying the degree of lung compromise in COVID-19 patients. The AI system has been experimentally deployed in the radiology department of the Brescia Public Hospital, Italy, since December 2020 during pandemic surges.

*Status:* We investigated the research question of what does "trustworthy AI" mean at the time of the COVID-19 pandemic. We started in April 2021 and conducted a Trustworthy AI self-assessment together with the key stakeholders, Alberto Signoroni (Department of Information Engineering), Davide Farina (Department of Medical and Surgical Specialties, Radiological Sciences, and Public Health) and their team. We have used the Z-inspection® process to assess the *ethical, technical, medical* and *legal* implications of using such an AI system at the Brescia Public hospital ("*ASST Spedali Civili di Brescia*").

At the time of assessment the AI system was deployed and used by main stakeholders.

We considered that it is crucial to perform a trustworthy AI assessment in urgent conditions, such as the COVID-19 pandemic. While it is clear that urgency requires a prompt reaction, it is not straightforward to define the tolerable ethical trade-off in place to accommodate such urgency. In fact, due to its nature, often one might be forced or led to pressure, for example, in deciding if the AI is mature or safe enough to be deployed in the hospital. We, therefore, recommend preventive assessment and early public adoption of AI ethical assessment to be ready for future emergency situations and other pandemics. Another point was the waiver from the Ethics Committee. Combining urgency and AI - lead an important ethics oversight function to be waived. How can we equip existing governance to deal with both AI and urgency?

The results of our assessment – completed in December 2021– have been submitted for publication (in revision)[52].

### 2.1.3 A *holistic* approach

Z-Inspection® uses a *holistic* approach, rather than monolithic and static ethical checklists. This is of great value when it comes to analyzing use cases, because it mirrors the complexity of the use context of an AI system. To be successful, the system must adequately meet technical, legal, ethical, and social requirements. Using a holistic process means that the Z-Inspection® assessment brings these different considerations together to provide an assessment of the system not only in different respects (technical, legal, ethical, social, etc.), but as a whole functioning socio-technical unit.

This interdisciplinarity is one of the most important aspects of our approach to ensure that a variety of expert methodologies, cultural ontologies, and disciplinary interpretations are expressed when assessing the trustworthiness of an AI system.

The complexity of the AI system and socio-technical deployment needs to be reflected in the structure and composition of the inspection team members.

Another strength of the holistic approach is that it is *dynamic*. A standard checklist does not adapt to the case at hand. The holistic approach on the other hand determines which issues are central for the use case at different stages of the process, and moves back and forth between intra- and inter-disciplinary discussions of which aspects of the case are most significant. In this way, the process has a certain degree of plasticity, which means that its assessments will be tailored to the use-case at hand. It also means that the assumptions that guide the AI system's creation and deployment – as well as the assumptions of those researchers conducting the Z-Inspection® – are exposed and evaluated in the course of the inspection. Finally, the dynamic lens reflects the commitment to an ongoing and iterative investigation of the harms and benefits of a particular AI system. While Z-inspection® is necessarily limited to a particular development phase, it provides space for participants to openly reflect and document what is known (and unknown) about the system's capabilities as a baseline for subsequent evaluations. As such, Z-inspection® is sympathetic to recent documentation work highlighting the dynamic effects of data-driven optimization techniques over time, such as "Reward Reports for Reinforcement Learning"[31].

### 2.1.4 The choice of experts

The choice of experts required for each use case has an ethical dimension since the quality of the analysis and the results depend on the diligent selection and quality of experts including them not being biased or in a position of conflict of interest. Domain experts may need to include several classes of expertise and practice, especially as a tool may impact the workflow of different categories of professionals.

Moreover, in case of a self-assessment of an AI system, the following are important considerations to be taken into account:

 - Special considerations should be taken into account of the potential behavioral bias of the stakeholders owing to the use case in the process of the evaluation.

-In those cases where the design/implementation and management of the AI system or parts of it is outsourced to a third party vendor: i) the third party vendor is not part of the self-assessment team to avoid possible conflict of interests, ii) the main stakeholders owing the use case need to declare of not having any conflict of interest with the third party vendor, and iii) they act as sole communication channel with the third party vendor. No direct communication between the rest of the evaluation team and the third party vendor is allowed.

*Practical Suggestion*

*An example of team composition is as follows:*

**Lead:** coordinates the process and the finalization of the interim issues report.

**Ethicist(s)**: help the other experts identify ethical tensions and dilemmas and how to solve these.

**Domain expert(s)**: (are ideally more than one to bring different viewpoints) assist inter alia in establishing whether there is a ground truth regarding the problem domain, and what this is.

**Legal expert(s) specialized for the specific domain**: due to being highly specialized in the field legal experts should be familiar with the problem domain area and/or have some understanding of the legal aspects of data protection and human rights.

**Technical expert(s)**: with specialty in Machine Learning, Deep Learning (including Neural network architectures) and data science.

**Representative of end users**: here it is important to identify who the "end users" are and have them involved in the assessment if possible.

**Representatives of those affected:** this could be patients or representatives of patients organizations for AI use cases in the domain of healthcare.

Preferable: **Social Scientists, Policy Makers, Communication specialists:** should ideally also be part of the assessment team.

*The role of Philosophers / Ethicists*

Philosophers / Ethicists should act as "advisors" to the rest of the team in order to assist team members with little ethics background in the interpretation of the seven requirements identified in the EU guidelines. In addition, they should be part of the process to identify ethical tensions between different understandings of the context of the use case, especially through the mobilization of the main ethical theories and approaches and be part of the mapping to the Trustworthy AI Framework. They should be available for ethics-related questions and guidance, including the links to values, which in the EU context include human dignity, freedom, equality, solidarity, democracy and human rights [51]. The understanding of these values, principles and legal requirements, are important. They should be available for identifying blind spots in perspectives and routines of Machine Learning practitioners. If they have use case-specific practical expertise (e.g. health / medical ethics) they could lead the part of the process that is to identify ethical tensions.

*Criteria for selection*

Team members should be selected based primarily on required skills and expertise – availability, motivation and interest in the case are essential but should not be the primary criteria for involvement. To ensure the quality of the inspection process, it is important that all team members respect specific areas of competency of each other. Later additions of experts to the team should be limited. It is preferable that later additions are avoided to keep the team's viewpoints balanced and the workflow of the team stable.

*Considerations for diversity in healthcare selection*

In the case of healthcare, there may be many different fields of expertise. Both as domain experts and as end users, the tool may impact workflow and patient care differently for different types of healthcare practitioners. Anticipated impact may also be perceived differently by researchers and public health specialists focusing on the same domain, adding additional insight. It is important to include different types of practitioners and other domain experts based on anticipated impacts of the tool, keeping in mind a healthcare worker may not use the tool directly but their provision of care may be impacted substantially.

In highly resourced settings there may be multiple specialties involved to ensure a good clinical outcome. In the case of cancer patients, one clinical issue may involve nurses, surgeons, oncologists, radiation oncologists, radiologists, laboratory technicians and microbiologists, and infectious disease specialists working together, along with others. On the other hand, in resource limited settings, fewer healthcare providers may be available, more may be done by individuals covering broad domains, with less specialized training, with time and resources quite stretched, and fewer diagnostics and care modalities possible. As such having a diversity of evaluators is important to foresee different concerns, though limiting the group size to manageable numbers of persons is still necessary.

However for the Z-inspection® process - expertise can also be drawn from outside the "resource limited settings" especially with technology as we learned during covid use case [52]

In addition, healthcare providers will have different concerns based on the context in which they work. Healthcare providers will work with different populations and may recognize different concerns regarding equity and diversity. Tools tested on certain populations may not be as effective in others and healthcare providers with different patient populations may recognize important gaps. There will be different medical-legal concerns regarding the tool, depending on the regulatory environment in which they practice, and this perceived legal climate reaches beyond formal regulations but also be determined by the risk group practices, hospitals, insurance agencies, and health systems are willing to take.

Healthcare providers may have different appreciations of the measurements of effectiveness of the tool. It will not just be accuracy that matters but sensitivity and specificity as well. Whether the tool would be useful as a screening tool or as a specific diagnostic tool will depend then on the landscape of other diagnostics and tools available as to whether a tool is needed and whether other tools are available as a safety net as needed.

> Healthcare workers may also recognize how a tool could disrupt current practice and this may require brainstorming about the downstream effects, which can be accomplished more effectively with a diverse group, with different insights.
>
> *Challenge*
>
> The main challenge is to make sure that all experts have a holistic view of the process and a good understanding of the use case. For that, all team members and relevant use case stakeholders need to be trained or train themselves on the EU regulation / Z-Inspection® process. This could also be done by reading documents backed by online training sessions and/or by producing "video-presentations" available to all the stakeholders related to the assessment.

### 2.1.5 Set-up: Definition of the boundaries

The set-up phase also includes the definition of the boundaries of the assessment, taking into account that we do not assess the AI system in isolation but rather consider the social-technical interconnection with the ecosystem(s) where the AI is developed and/or deployed.

Taking inspiration from work by Dobbe et al. [32], some of the most important ethical and political considerations of AI development rest on the decision to include or exclude parts of the context in which the system will operate. Z-Inspection® addresses this by making the boundaries of the assessment explicit, articulating which features are included or not and why in relation both to abstract ethical concerns and the concrete details of the system's operation. At the same time, given the inherent diversity of populations and the need for equity, particularly for medical devices, it is important to expand the assessment as far as feasible, while recognizing and clearly stating any limitations.

Another aspect that is relevant here, is that when we consider what are the boundaries for the assessment, we may consider two requirements that we have added to the seven requirements for Trustworthy AI, namely [11]:

-Assessing if the ecosystems where the AI is developed and/or deployed respect values of Western European democracy, and

- Assessing if the use of the AI results in concentration of power.

Therefore by taking into account these two requirements, the definition of the boundaries for the assessment becomes an ethical imperative with deep influence on the potential results of the assessment.

## 2.2 Socio-technical Scenarios to identify *issues*

One of the specific aspects of the methodology involves using what we call socio-technical scenarios [33], in order to anticipate possible uses and problems of the system under review.

We conduct self-assessments together with the stakeholders owning the use cases i.e., the ones in charge of the design, development, deployment, or evaluation of the AI system in question. In the rest of the paper, we will use the term *stakeholders* to denote the actors who have direct ownership of the development and deployment of the AI system.

The scenarios are built around possible everyday life experiences that could result from the use. Anticipating those experiences highlights possible problems that could arise from such use. One can then 'look' at the situation from different points of view, highlighting different approaches and appraisal of the technology at hand. The team draws from their diverse experiences in technological assessments to debate the specific context of the situation. This prevents abstract oppositions between general principles [34].

This method proved to be useful in allowing lively discussion of specific problems related to the use case while engaging at the same time tensions between fundamental principles and ultimately various understandings of our moral obligations (understandings that can be linked to the main ethical theories).

By collecting relevant resources, a team of interdisciplinary experts creates socio-technical scenarios and analyzes them to describe the aim of the AI systems, the actors and their expectations and interactions, the process where the AI systems are used, and the technology and the context (*ecosystem*).

The scenarios themselves are created based on the perceived ethical stakes of the system's specification (including its relationship with the prior socio-technical context), how its operation is intended to affect the interests of various groups, and how the terms of its deployment reflect or enact a consensus across stakeholders for its appropriate use. In past cases, different Z-Inspection® teams exercised discretion in how these scenarios were created and internally discussed in the process of carrying out the inspection. The outcome of this first step is the identification of a number of issues to be assessed.

## 2.2.1 How to start?

To start with, the team of experts meet together with the stakeholders owning the use case in a number of workshops (possible also via video conference) to define socio-technical scenarios of the use of the AI systems.

## 2.2.2 Small vs. Large Team of Experts

The size of the team is correlated to the complexity of the AI assessment. We have been working on different use cases with different team sizes.

In one use case, we had a large team including over 40 interdisciplinary experts and we had to split the work in parallel working groups.

In other use cases we did not have to split the work in parallel working groups since we had a midsize team including around 20 interdisciplinary experts.

There are pros and cons for this decision.

If the team is too small and it does not reflect the true interdisciplinary nature of the assessment work, the assessment work will likely be incomplete. If the team is too big with too much overlap of knowledge and expertise, the assessment process may become cumbersome and delayed.

*Practical Suggestion*

> Lessons learned from working with the use cases were that low to medium teams with a limited number of participants, but including domain experts from all categories, proved to be the most efficient.
>
> Suggested team size for a low-medium complexity use case:
>
> (min 10-up to 20 experts) for the first use case.
>
> e.g.:
>
>     1 Lead
>     2 Ethicists
>     2 Legal
>     3 Technical
>     3 Domain
>     1 Representative of end users
>     2 Stakeholders owning the use case

> There may be some overlaps in categories, especially when considering domain experts and end user representatives. Given that some settings may have different types of domains involved, one may need a larger variety of domain experts to ensure the topic is covered. For example, in the healthcare domain one use case might include several clinical fields from ophthalmology to neurology to radiology to infectious diseases.
>
> The most important aspect is to include specialists from all fields e.g. law, IT/tech, philosophy, ethics, medicine. A larger span of domains provides various advantages since the team can draw conclusions and analogy on a broader spectrum of real world similar use-cases or problems.

### 2.2.3 Infrastructure

Meetings, sharing of information, and write-ups can all be virtual. Virtual meetings can be conducted on online platforms, content can be shared and written on shared docs, and communications can be continued through group emails. Recordings of meetings with transcribed discussions and preserved message discussions can be the raw data for the Z-Inspection® log protocol (documentation). The recorded meeting can then be shared and serve as a template for further discussion and as a knowledge base. For more structured access to information, we linked the relevant pieces of the document to the corresponding part of the recording so it was easier for participants to verify statements in the document or check for additional context. More information on the document template and how to connect it to recordings are available in the appendix. Such living documents are easily accessible to all, but can become unstructured and improvements in this process are still needed. The document served as a catalog of questions to the stakeholders for further refinement and clarification.

One lesson learned was the importance of getting the "facts" about the AI system, the context and the implementation clarified early. Some important questions remained unanswered for some time, making the identification of "issues" somewhat speculative.

E-mail groups allowed a wide variety of members from different groups to be involved but email threads proved suboptimal as unmoderated discussion threads split. We are considering comparing different forum-based communication systems in order to see which one is the most effective for this kind of collaborative research work.

## 2.2.4 Creating Socio-Technical Scenarios

To start a new assessment process, we hold meetings with stakeholders, where they tell us about the AI system. They give us some context on where the system is used, what problem it is intended to solve (goals, objectives), what steps they have taken to ensure the system is solving the problem, and how the system is currently used.

We found it useful to collect a written summary of this meeting, where the information is organized according to the following structure:

1. **Aim of the system**
   Goal of the system, context, why it is used.
2. **Actors** (*primary:* directly involved with the use of the AI system, and *secondary* and *tertiary* only indirectly involved with the use of the AI system)
   Who designed and implemented the system? Who has authorized the deployment of the system? Who is currently using the system? Who are the end users for this system? Who is directly influenced by decisions made by the system? Who is indirectly influenced by decisions made by the system? Who is responsible for this system?
3. **Actors' Expectation and Motivation**
   Why would the different groups of actors want the system? What are their expectations towards the system behavior? What benefits are they expecting from using the system?
4. **Actors' Concerns and Worries**
   What problems / challenges can the actors foresee? Do they have concerns regarding the use of the system? What risks are they concerned about with the system? Are there any conflicts?
5. **Context where the AI system is used**
   What additional context information about the situation where the AI system is used? (e.g. urgency, budget constraints, for profit, academic, conflicts, environmental). What are potential future usage of the AI system?
6. **Interaction with the AI system**
   What is the intended interaction between the system and its users? If and how the 'human in control' aspect is envisaged? Why is it like this?
7. **AI Technology used**
   Technical description of the AI system. An important part of considering AI trustworthiness is that it is robust and if the technical description is not clear, this cannot be assessed.

8. **Clinical studies /Field tests**

    Was the system's performance validated in (clinical/field tests) studies? What were the results of these studies? Are the results openly available?

9. **Intellectual Property**

    What parts of the AI system are open access (if any)? What IP regulations need to be considered when assessing / disseminating the system? Does it contain confidential information that must not be published? What is and how to handle the IP of the AI and of the part of the system to be examined. Identify possible restrictions to the Inspection process, in this case, assess the consequences (if any) Define if and when Code Reviews are needed/possible.

10. **Legal framework**

    What is the legal framework for use of the system? What special regulations apply? What are the data protection issues?", "Was the data aspect compliant with the GDPR?

11. **Ethics oversight and/or approval**

    Has the AI system already undergone some kind of ethical assessment or other approval? If not - why not? If so, was this internal/external, volunteer/regulated, what was covered? One could argue that if they already had some, it is unlikely they will engage with Z-inspection®, but from asking the question - we also get an understanding of the gap(s). Did they get a waiver? Was there a clearing, but it was very light or internal and not considered sufficient?

With the information organized this way, people are encouraged to go through the materials again and ask questions in the document where they need additional clarifications from the stakeholders or developers. This helps bring everyone to a shared understanding of the system which helps create socio-technical scenarios based on possible uses. It also helps stakeholders to identify where they need to provide additional information on the system. For this part of the process, we developed a toolkit as indicated in the Appendix.

Upfront it is important to agree on a time frame for inspection with set dates. Meetings should be agreed upon with the final inspection team.

## 2.3 Develop an evidence base

Technology is generally designed for a highly specific purpose, however, it is not always clear what the technologies unintended harm might be. Therefore, an important part of our assessment process is to build an evidence base through the socio-technical scenarios to identify tensions as potential ethical issues to be discussed further.

**Claims** for technological capability (for example aim, performance, architecture, or functionality, etc. ) serve as an important input in developing the **evidence base.** This is an iterative process among experts of the assessment team with different skills and backgrounds with a goal to understand technological capabilities and limitations [35].

*Practical Suggestion*

> The inspection team needs to have a shared and informed understanding of the use case to work with. The "use case owners'' are needed to provide the required information.
>
> It is important that all team members have a clear understanding of which information (including claims) is provided by the use case owner and which information (analysis/results etc.) was generated by the inspection team; clear separation of the presentation of both kinds of material can help. This separation should be clear also in the final inspection document.
>
> The use case owners should be responsible for the use case presentation section of the working document. Answers by the use case owners to questions from the inspection team during the inspection process should be organized in a systematic Q&A part of the working document. The use case owners must also take responsibility for the content of this Q&A part.
>
> Our experience has shown that a missing separation and content control by the use case owner can result in misinformation or even speculation on the side of the inspection team members. This can not only extend the inspection process and frustrate team members, but even harm the inspection results. These negative effects of an unclear and uncontrolled presentation of information, analysis and results increase with the size of the inspection team. This is a difficult tension. However, we have also seen the opposite tendency for use case owners to control a narrative on how a tool would be used in the real world, including instances where use case owners have not had experience with the larger downstream effects such tools can have, especially as tools in the hands of users may be used in unexpected ways and that is the intention of these groups of experts to tease apart.
>
> While the information coming from the use case owners is absolutely crucial to the process, our experience has shown that the assessments must not be one of the intentions of the use case owners. This is why the process should avoid being an examination of the use case owners. The claims, arguments and evidence (CAE) framework can help with the structuring of the use case in a clear and precise form that is supported by evidence. For example, each of the claims should be about only one specific property of the system and at the same time, it should be phrased in a way that is clearly verifiable or falsifiable. The CAE framework also provides guidance on how to disseminate complex claims into easier ones [36].

The use case owners are "responsible" for providing thorough **evidence and arguments for their claims.** Speculations might compromise the assessment if the use case is not presented in a clear and precise form with evidence provided because the discussion becomes general and not use case-specific.

A person needs to be in charge of managing Q&A and organizing answers and evidence so that it is less time intensive to check what is a claim, what is evidence, and what argument.

Structuring information in this way helps the team members, as it gives them a better idea of the actual capabilities of the system, as well as routes for verification of these capabilities.

It also helps the owners by showing them where the information they provide can be misunderstood and what kinds of evidence they need to produce as well as a way to assess if the evidence they produce is sufficient.

A Q&A log should be started and a person should be responsible for updating this documentation and for "quality control". Furthermore, it contains a glossary of otherwise ambiguous terms with project-specific definitions. It is organized to separate concerns and allow for clear responsibility of an expert group. The log should have thematic subject headers: for easier indexing and finding information e.g., data, algorithm / ML, medical, legal, ethical… The answers should always be provided with evidence / argumentation to allow for verification of the answers, e.g., concerning consistency or process-compliance.

*Challenge*:

If the use case is not presented in a clear and precise form with evidence provided, speculation comes in and "deteriorates" the results – i.e. discussion becomes general.

We suggest building a solid knowledge / evidence base among all team members of the use case before the inspection starts and also a solid Q&A log during the inspection process.

Experts may approach the use case quite differently:

- Interpretations of and expectations for the AI tool being inspected may differ
- Focus of interest may be very different
- Certain areas of the AI tool, whether algorithmic or domain-specific, may be overlooked by those with a different area of focus

> - Some may foresee likely misuse or unintended consequences of the tool, whereas others may focus on the intended use and results
> - Those who are domain experts or end users may focus on more pragmatic areas, such as brainstorming actual practical use or trying to understand the underlying reasons for its effectiveness - and hence also potential pitfalls
> - Experts in different fields will see the tool quite differently. What may be considered a lack of knowledge can just be a different lens. It's crucial the team understands that there will be very different perspectives based on the specific role or subdomain different experts represent.

## 2.3.1 Managing Different ViewPoints

Managing different viewpoints between experts composing the assessment team is an essential part of the process.

One of the key lessons learned is that there may be tensions when considering what the relevant existing evidence to support a claim is.

When working in co-creation, as for example in the case of the skin cancer detection AI tool [25], there were *tensions* between the various arguments linking evidence to support the choice of a design decision derived from the different viewpoints expressed by domain experts.

We show below and example as such *tensions* when we co-created the skin cancer detection AI tool [25]:

*Argument* **:** Malignant melanoma is a very heterogeneous tumor with a clinical course that is very difficult to predict. To date, there are no reliable biomarkers that predict prognosis with certainty. Therefore, there exist subgroups of melanoma patients with different risks for metastasization, some might never metastasize and diagnosing them would be overdiagnosed.

*View Point*: Dermatologist.
Early detection of malignant melanoma is critical, as the risk of metastasis with worse prognosis increases the longer melanoma remains untreated.

*View Point:* Evidence Based Medicine Professional.
There are no reliable biomarkers that can predict the prognosis of melanoma before excision. There are patients who survive their localized melanoma without therapy. Therefore, the early diagnosis does not necessarily mean a better prognosis; on the contrary, there is a risk of poor patient care due to overdiagnosis.

For the skin cancer detection AI tool [25], by considering such various viewpoints together with the AI engineers, who originally designed the first prototype, it was possible to re-evaluate the aim of the system.

During the co-creation process, the discussion with different experts in our team, including experts in public health and healthcare, among them dermatologists and specialists in evidence-based diagnosis, ethics, law, and ML, prompted the main stakeholder and owner of the use case, the team of German Research Center for Artificial Intelligence (DFKI), to redefine their stated main aim of the system. When evaluating the use case design, it soon became clear that different stakeholders have different scopes, timeframes, and the population in mind. Thanks to the heterogeneity of our team, such differences and tensions were confronted. For each viewpoint, intensive exchange and communication took place between various domain experts, with different knowledge and backgrounds, all of whom form part of our team.

Managing different viewpoints also made evident that the process requires researchers to bring their own ideas and arguments to the discussion of aspects of the case while at the same time understanding that their input is a contribution to teamwork, not a matter of "winning the argument."

*Practical Suggestion*

- Different Viewpoints among Domain Experts should be seen as an opportunity to, on the one hand, bring to the discussion one's ideas and points of view on aspects of the case and, on the other hand, seeing it as input to teamwork, not as a matter of "winning the argument." There likely is no right or wrong answer. Tools are not binary good or bad, but have good qualities and reasons for concern. Sometimes simply recognizing the concerns can make the tools more effective in practice. This could also lead to recommendation from the inspection on how the system owner can mitigate a risk or manage concerns in practice. (Eg. better training of some groups of users. Implement a protocol for use or clarify to patients how data is used and ask consent, etc.)

Opposite points of view may both have good arguments in favor of them, and it might not be straightforward if one of them is right or wrong, but it is very useful to be aware and articulate them both.

Who is "qualified" to give strong evidence?  We could introduce different levels of what constitutes "evidence".  Strong evidence is when testing is possible. However, testing is not always possible. We look at peer-reviewed journal articles supporting a claim. This is also evidence. When domain experts have different viewpoints, then we list such different viewpoints and related supporting evidence as tensions. This was done in this co-design use case [25].  However "evidence" is somewhat complicated in science. Testing supports claims until someone finds support for another claim that might contradict or expand on the first claim. That happens all the time and is how science evolves (i.e. evolution, physics, medicine). Strong support for claims is the best we have.

Another point to consider, case by case, is when special supporting opinions by some experts carry a heavier weight and they should be considered as a "threshold" for the assessment.

When possible create an Institutional Review Board (IRB)/Ethical Review Boards to discuss how we assess ethical questions differently, and how these reviews can complement and impact each other in the long term. Some IRB reviews have been done including different end users than our groups would have imagined, as we are thinking of broader downstream impacts. (Some have looked at end users as clinicians, rather than patients, resulting in very different IRB reviews).

It might also be useful to create a panel of end users to review on a regular basis the evaluation process and provide feedback.

## 2.4 How to describe "issues": Using free text and an open vocabulary

In the case of parallel Working Groups(WGs), each WG analyzes the socio-technical scenarios and produces preliminary reports - working independently and in parallel to avoid cognitive biases and take advantage of their unique perspective and expertise.

Preliminary reports are shared with the entire team for feedback and comments.  These interdisciplinary interactions among experts with different backgrounds allow each WG to consider the viewpoints of other experts when delivering their final reports.

Each final report is then written using free text and an open vocabulary to describe the possible risks and issues found when analyzing the AI system. For example, each WG report may list the identified ethical, technical, domain-specific (i.e. medical) and legal issues described using an open vocabulary.

When the team size is relatively small (e.g. less than 20 people), then no WG is created and the entire team carries on the assessment. In this case, the final report is the result of the work of the assessment team.

We present below an example of an *issue* identified when analyzing the emergency tool to detect cardiac arrest tool [15]:

---

**ID Ethical Issue: E4, Fairness in the Training Data**

*Description*
The training data is likely not sufficient to account for relevant differences in languages, accents, and voice patterns, potentially generating unfair outcomes.

**Narrative Response** (*Open Vocabulary*)

There is likely empirical bias since the tool was developed in a predominantly white Danish patient group. It is unclear how the tool would perform in patients with accents, different ages, sex, and other specific subgroups. There is also a concern that this tool is not evaluated for fairness with respect to outcomes in a variety of populations. Given the reliance on transcripts, non-native speakers of Danish may not have the same outcome. It was reported that Swedish and English speakers were well represented but would need to ensure a broad training set. It would also be important to see if analyses show any bias in results regarding age, gender, race, nationality, and other sub-groups. The concern is that the training data may not have a diverse enough representation.

---

*Practical Suggestion*

Important questions should be raised by the inspection team members and systematically organized as early as possible. It is the task of the WG leader and the overall project leader to make sure that questions are categorized by importance and that supporting evidence is being provided accordingly.

Try to avoid that evidence for some important questions is provided by stakeholder team too late – i.e. only at the end of the process

*Challenge*:

> Expert teams may uncover very different issues than the overall team may expect. For example, those in clinical medicine may have a very different practice than those outside of medicine recognize. Because expert teams may uncover very different issues than the overall team may expect, it is important to facilitate the discovery of important issues as early as possible in the process, e.g., by ensuring quick input from expert teams.

### 2.4.1 Concept Building in Practice

Whittlestone et al. [35] introduce the notion of Concept Building, by observing that an important obstacle to progress on the ethical and societal issues raised by the use of AI is the ambiguity of many central concepts currently used to identify salient issues [35], [37].

In part, because the successful operation of a given AI system requires precise knowledge of how it is to be used and what relationship its behavior will have to human settings, commitments to "fairness" or "safety" may have to be worked out at different stages of the ML pipeline and life cycle. This ambiguity or indeterminacy may take several forms. First, there are no clear rules on who audits whom and at what intervals. Secondly, auditors might lack relevant information for sound auditing proceedings. Thirdly, auditing measures are susceptible to adversarial behavior, meaning attempts to game auditing metrics. Fourthly, auditing may miss investigating the whole range of AI machine behavior in real-world settings. Fifthly, non-recurrent audits are not compatible with agile, fast-paced technology development cycles. Sixthly, conceptual constraints concern potential disagreements on ethical principles, normative values, dilemma resolving, etc. Apart from that, AI audits require interdisciplinary research settings, in which finding common vocabularies and perspectives can be challenging. Whittlestone et al. [35] suggest the use of Concept Building to manage terminological overlaps, differences between disciplines, and differences across cultures and publics.

They mentioned that conceptual complexity includes mapping and clarifying ambiguities; bridging disciplines, sectors, publics and cultures; building consensus and managing disagreements.

We have been using Concept Building in practice. This will be explained in Sect. 2.5 below.

## 2.5 Identify value conflicts and trade-offs

If the approach suggested by the AI HLEG Ethics Guidelines for Trustworthy AI [13]may at first hand be seen in line with principlism, (which links back to Beauchamp and Childress and their very influential book "Principles of Biomedical Ethics"[38]), the status of the guidelines and of the Z-inspection® initiative should be compared with this approach. Beauchamp and Childress discuss four central ethical principles in healthcare contexts: beneficence, non-maleficence, respect for autonomy, and justice. According to their approach, there is no clear hierarchy among these ethical principles, but their roles have to be balanced according to their respective context [38], [39]. While this has the advantage of flexibility, the downside is that it may not always be clear how to balance various principles. The principlism approach to bioethics has been used for more than 30 years, but it has also been contested, especially for its reliance on abstract 'principles' that are not really guides for action [40], [41]. Many authors have been seeking a less abstract interpretation of these principles in order to understand them in the context of healthcare [42].

While such a framework has been used in healthcare contexts, only rarely have there been case studies that directly seek to analyze the implications of the various principles in detail in the context of artificial intelligence, a notable exception being the work done in [43]. In those case studies, the assessment of the roles and weight of the various principles – ethical pillars and requirements – turned out to be a field that needs further investigation. While the group was able to identify tensions between the various ethical pillars, requirements and sub-requirements, the very question of how to balance the pillars, requirements and sub-requirements turned out to be very complex.

The four central ethical pillars defined by the *Ethics Guidelines for Trustworthy AI* [13], respect for human autonomy, prevention of harm, fairness, and explicability, bear considerable similarity with the framework developed by Beauchamp and Childress, except that beneficence has been exchanged for explicability. This is in line with what others have suggested as central principles for ethical AI [3], [38].

What makes the ethical evaluation more complex, however, is that each of these ethical pillars contains seven requirements from which to choose: Human Agency and Oversight; Technical Robustness and Safety; Privacy and Data Governance; Transparency; Diversity, Non-Discrimination and Fairness; Societal and Environmental Well-Being, and Accountability. Each of the requirements comes with several sub-requirements.

The requirement level also contains options that are conceptually in direct relation to the central ethical pillars. For example, "Human agency and Oversight" or "Privacy" can be seen as aspects of "Respect for autonomy", and "Safety" or "Societal and Environmental Well-Being" as intertwined with "Prevention of Harm".

What on the one hand can be considered an advantage is that the pillars - requirement - sub-requirement framework offers additional aspects to consider in the ethical assessment, but has the clear downside that the three-level assessment comes with three levels of decision-making. The first-level decision (i.e. pillar) clearly influences the decisions further down the line (i.e. requirement and sub-requirement).

To give an example, a situation where the team identifies an ethical tension between the central ethical principles of respect for autonomy, prevention of harm and explicability/transparency: If the team decides to begin with the ethical pillar "Respect for Autonomy", the plausible requirement and sub-requirement options will be different than if the team begins with the pillar "Explicability".

Furthermore, some concepts, such as privacy, could be considered an aspect of respect for autonomy or of prevention of harm, which increases the assessment options and further complicates the assessment process.

It has to be emphasized that the group did not discuss the various values and ethical principles and their balancing in an abstract, high-level way, but referred to the low-level technological requirements of the use case, using the mid-level guidances offered by the EU guidelines (especially the 7 requirements) and then discussing the roles of these values and principles in the respective context. This can clearly be seen as an attempt to learn from previous applied ethics debates on the role and use of principles, especially, in the history of bioethics and biomedical ethics, in order to avoid a strictly principlism approach to such technological assessment.

Overall, during the assessment process, conflicts or tensions arose between ethical principles and values, for which, in theory as well as in practice, there is no fixed solution. This explains why "In line with the EU fundamental commitment to democratic engagement, due process and open political participation, *methods of accountable deliberation to deal with such tensions should be established*" [13].

In evaluating specific use cases, the group had to come up with a process that serves to unify the assessment process and allows agreement between the various experts with different backgrounds. The process begins with describing the tensions between ethical values using open vocabulary, and gradually narrowing the options down to finally agree on a close vocabulary description. The close vocabulary is from the EU framework. As mentioned above, between the levels of ethical pillars, requirements and sub-requirements, the level of requirements turned out to be particularly useful in that it provides an adequate level of granularity as described in Section 2.6.

## 2.5.1 Ethical Tensions

In view of the complex conflicts between the various pillars, requirements and sub-requirements and the obvious need for trade-offs resulting from these conflicts, the group decided to recur to a pragmatic process.

In this, we used the umbrella term "tension" to refer to various ways in which norms, values and principles can be in conflict. With this term, we denote tensions between the pursuit of different values in technological applications rather than abstract conflicts between the values themselves.

This is clearly related to our methodological choice to assess technological systems using socio-technical scenarios. We are not looking for abstract tensions between general principles, but situations where highly valued principles are in conflict in a specific technological setting. Of particular relevance are conflicts where norms, values or principles are mutually exclusive and thus cannot all be materialized or not all be materialized at the same level or with the same priority. The ethical tension is most emphatically embedded in a technological device when highly praised norms or values are in conflict in the device itself or in its social use, and when a choice (at the design stage, or the deployment stage, or the use assessment stage) must be made. The whole methodology of the Z-inspection® leans towards such an ethical assessment of AI systems.

After having reflected on the various norms, values and ethical principles that play a role in the technological application, the next step in the assessment process consists in specifying the most relevant ethical tensions. In general, at this step, central tensions between two or more relevant aspects were identified, with the focus on tensions between two norms, values or principles. The task is to describe these tensions in open language. For example, in a medical use case, examples could be: a tension between quality of services and autonomy; or between upholding standards and prevention of harm; or between efficiency and autonomy.

In order to facilitate the process of identifying tensions, we recurred to the catalog of tensions defined by Whittlestone et al. [35]. This list of suggestions made it easier for team members to categorize a specific problem discussed in the context of a technology application as relating to a specific norm, value or principle, such as, for example, accuracy, fairness, personalization or solidarity. And it made it easier to identify and characterize specific tensions.

This catalog proved to be very useful in the assessment process, even though in itself it is not enough to resolve the complex assessment that results from several levels and more than two or three ethically relevant aspects. (See below, section 2.6.)

Practical Suggestion

> Since some experts in the assessment team may not have a background in ethics, we use a predefined catalog of ethical tensions as examples to help the identification of "issues".
>
> This is another positive aspect of using socio-technical scenarios, since participants do not need to be fully trained in theoretical ethics to be involved in the debate. Ethical experts in the group can provide some expertise regarding the theoretical aspects of the discussion, while participating in the assessments of the socio-technical scenarios.
>
> Concretely, we use the catalog of tensions defined by Whittlestone et al. [35], namely:
>
> *Accuracy vs. Fairness*
> *Accuracy vs. Explainability*
> *Privacy vs. Transparency*
> *Quality of Services vs. Privacy*
> *Personalisation vs. Solidarity*
> *Convenience vs. Dignity*
> *Efficiency vs. Safety and Sustainability*
> *Satisfaction of preferences vs. Equality*
>
> And we consider also their proposal on how to identify **further tensions**
>
> *"Thinking about tensions could also be enhanced by systematically considering different ways that tensions are likely to arise.*
>
> Whittlestone et al. [35]: points to three axis for such considerations: power, time and locus as follows:
>
> ***"Winners versus losers***. *Tensions sometimes arise because the costs and benefits of ADA-based technologies are unequally distributed across different groups and communities.*
>
> ***Short term versus long term***. *Tensions can arise because values or opportunities that can be enhanced by ADA-based technologies in the short term may compromise other values in the long term.*
>
> ***Local versus global***. *Tensions may arise when applications that are defensible from a narrow or individualistic view produce negative externalities, exacerbating existing collective action problems or creating new ones. "*
>
> This helps to flag the broader issues linked to *winners* versus *losers* i.e. power, *Short/Long term* i.e. time and *Local versus Global* - i.e. the locus or scope - or the "good for whom?" question.

## 2.5.2 Classification of Ethical Tensions

Once the ethical tensions have been identified as part of the case study assessment, the next question is how – if at all – these ethical tensions can be resolved. Thus, the next step in the assessment process consists in deciding which of the options available to choose. For example, in the case of an identified ethical tension between efficiency and autonomy, whether to choose the option that respects autonomy or the option that safeguards efficiency.

In this step of the assessment, the distinction between true dilemmas, dilemmas in practice, and false dilemmas, as suggested by Whittlestone et al. [35] proved to be very useful, as we have found in group sections it can be the area that is the least understood/known to those in domain area groups.

In general, dilemmatic situations are situations where a difficult choice between two options has to be made. There are different types of dilemmas, however.

Genuine ethical dilemmas can be characterized as situations where a choice has to be made between two or more options, but, no matter how the decision is made, there will be negative moral consequences. Sinnott-Armstrong gives the following definition for a genuine moral dilemma [44]:

"a situation where an agent has a *strong* moral obligation or requirement to adopt each of two alternatives, and neither is overridden, but the agent cannot adopt both alternatives."

In genuine ethical dilemmas, there are strong reasons to do each of two things, but only one can be done. Dilemmatic situations involve a conflict of two norms. When a decision is made and one option chosen, no matter what the decision is, the option chosen will conflict with one of the norms, values or principles.

Thus, with an ethical tension, for example, an ethical tension between efficiency and autonomy, a "true dilemma" implies that the group can either choose an option that is in conflict with autonomy, or an option that is in conflict with efficiency. There is no "good" option available that satisfies both norms, values or principles.

The interdisciplinary approach in the Z-inspection® can help surface such tensions from different perspectives. To anticipate and prioritize among principles and values of "true dilemmas" a similar interdisciplinary dialogue approach should be considered prior to determining whether and how to apply an AI use-case. National Ethics Committees or committees, if provided with the skills and capacity, could have a role in ensuring that societal stakeholders have a voice in determining which values to prioritize.

In contrast, a "dilemma in practice" is a dilemma where an ethical tension exists, but where this tension can be overcome in principle. The tension exists only for practical reasons and could be overcome, for example, if more money would be available or if a different technological approach would be chosen.

A "false dilemma" is a situation where two options with conflicting norms, values, and principles exist, but where it is not the case that a forced choice between these two options has to be made. The reason is that another, third option exists that is less dilemmatic, i.e. that does not require a choice between two important norms, values and principles. In the example with an ethical tension between efficiency and autonomy, this could be an option that is efficient but does not risk autonomy because it makes an additional level of data protection available.

*Practical Suggestion*

> Once ethical tensions have been identified, participants should try to classify the ethical tensions in 3 categories as suggested by [35] in order to assess the possible solutions to these tensions. Are the tensions identified exposing:
>
> A **True dilemma**, i.e. a conflict between two or more duties, obligations, or values, both of which an agent would ordinarily have reason to pursue but cannot; a
>
> A **Dilemma in practice**, i.e. "a tension which exists not inherently, but due to current technological capabilities and constraints, including the time and resources available for finding a solution"; or a
>
> A **False dilemma**, i.e. "a situation where there exists a third set of options beyond having to choose between two important values".
>
> This should be an integral part of the discussion surrounding the ethical tensions.

We present below an example of a *tension* identified when analyzing the emergency tool to detect cardiac arrest tool [15]:

| **ID Ethical Tension (Open Vocabulary): ET4** |
| --- |

> Kind of tension: **True dilemma.**
>
> Trade-off: *Fairness vs. Accuracy.*
>
> Description: The algorithm is accurate on average but may systematically discriminate against specific minorities of callers and/or dispatchers due to ethnic and gender bias in the training data.

## 2.6 A consensus process based on mapping: Open to Closed vocabulary

To be able to consolidate an assessment process, which comprises contributors with different backgrounds and expertise, we designed a process that would facilitate an agreement between participants. To this end, the team of interdisciplinary experts firstly map issues by freely describing the issues, using open vocabulary, and then use closed vocabulary to assign these issues to the 4 ethical principles and the 7 requirements for trustworthy AI.

We rank the *mapped issues* by relevance depending on the context. (e.g. Transparency, Fairness, Accountability.)

The "*issues*" described in free text are then mapped by each WG using templates (called rubrics) to some of the four ethical principles and the seven requirements defined in the EU framework for trustworthy AI [13].

With this mapping, the reports are developed **from an open vocabulary to a closed vocabulary** (i.e. the templates). We call this process "*mappings*". Each working group worked independently from each other, and adopted different/similar strategies to perform the *mappings*. The *mappings* and the common vocabulary they allow to build are especially important in the interdisciplinary context that any technological assessment necessarily implies.

The *mappings* require participants to translate their own disciplinary methods and cultural perspectives into a single language that everyone speaks. This entails a commitment from the listener to highlight confusions or ambiguities as they arise, and a commitment from the speaker to pause and clarify before moving forward. While challenging, this allows each participant in the Z-inspection® process to fully participate towards creating the closed vocabulary mappings, while contributing with their own expertise to the analysis.

We present below an example of an *mapping* identified when analyzing the emergency tool to detect cardiac arrest tool [15]:

> **ID Ethical Issue: E4, Fairness in the Training Data**
> *Description*
> The training data is likely not sufficient to account for relevant differences in languages, accents, and voice patterns, potentially generating unfair outcomes.
>
> **Narrative Response** (*Open Vocabulary*) There is likely empirical bias since the tool was developed in a predominantly white Danish patient group. It is unclear how the tool would perform in patients with accents, different ages, sex, and other specific subgroups. There is also a concern that this tool is not evaluated for fairness with respect to outcomes in a variety of populations. Given the reliance on transcripts, non-native speakers of Danish may not have the same outcome. It was reported that Swedish and English speakers were well represented but would need to ensure a broad training set. It would also be important to see if analyses show any bias in results regarding age, gender, race, nationality, and other sub-groups. The concern is that the training data may not have a diverse enough representation.
>
> **Map to Ethical Pillars/Requirements/Sub-Requirements** (*Closed Vocabulary*)
>
> **Fairness** > *Diversity, Non-Discrimination and Fairness > Avoidance of Unfair Bias.*

## 2.6.1 Mapping Strategy

An important lesson from the mapping process was that a mapping in terms of the four principles of Trustworthy AI turned out to be too coarse, whereas deploying the sub-requirements presented the group with a multitude of options that made the mapping too difficult. As it turned out, focusing the mapping on the seven requirements proved to be a useful conceptual middle ground for the mapping process.

Example of a Mapping Strategy

In one working group (*WG Ethics and Healthcare*), the mapping of issues identified in the WG report was organized using the following process:

At the *initial meeting*, they made a list of the key issues that they found to be present in the WG report. The list merely stated keywords, with no description of the issues. They then divided the issues between them and each member of the group made a description of his/her selection of the issues. The descriptions formed the basis of a second meeting at which they initiated the mapping of issues to ethical pillars, requirements and sub-requirements.

At the *second meeting*, they discussed the mapping of a couple of the issues identified. This involved quite a bit of clarification and discussion of their understanding of the pillars and requirements. Moreover, the group did not get around to mapping all the issues to the pillars, requirements, and sub-requirements. However, the meeting provided an extremely useful lesson in how differently experts, even experts with very similar backgrounds (in this case bioethics), may understand and apply these notions. Thus the meeting served to ensure that the group had a shared interpretation of the closed vocabulary. In turn, this meant that the discussion shaped and structured the way group members came to understand the issues.

On the basis of doing one mapping together at the end of the meeting, they decided that they would each map the issues they had described and then meet and discuss these suggested mappings.

They did this at the *third meeting*. At this point, the group was in a position to reach a consensus about the mapping of the issues expressing a common understanding.

### 2.6.2 Challenges of Mapping

We found that the mapping of an issue is often debatable and strongly depends on the background of the person performing the mapping. Disagreements regarding the mappings within the groups were resolved by group consensus. Thus, in one use case [52], across the different working groups, the whole team identified a large number of issues (over 50) that needed further consolidation.

An important lesson from the mapping sessions is that it is often not obvious which of the pillars or requirements applies to an issue. In many cases, multiple pillars or requirements can apply and a decision must be made about which one is the most applicable.

On the positive side, the nature of the Z-Inspection® process allows us to overcome this obstacle and come to an interdisciplinary consensus.

Based on the lessons learned in the Z-Inspection®, an approach which is not interdisciplinary risks having a one-sided perspective on the principles and requirements, and will be less likely to identify and solve ethical issues.

## 2.6.3 Consolidation of Mappings

The next step was to consolidate the mappings produced by the various WGs into a consistent list. For this task, we created a dedicated working group (coined *the Mappers*). The Mappers were to group the issues that had been mapped to the same requirements of the EU framework for trustworthy AI.

The consolidated lists of WG issues for each of the seven requirements were reviewed so commonalities and differences could be identified and discussed before final consolidation.

The method highlighted how different perspectives could lead to similar issues being mapped to different requirements.

An important initial lesson learned by the Mappers was that it was challenging to decide on a method for how to approach the fact that different WGs had similar issues mapped to different requirements. This required the Mappers to develop a more detailed strategy for how to do the mapping.

### 2.6.3.1 Example of consolidation Mapping strategies

The initial meeting of the Mappers made clear that the mapping process would require several steps in order to result in a valid consolidated list due to the large number of identified issues the consolidation was performed in two steps.

The meetings of Mappers also made clear that mapping the issues to the 4 principles alone would be too high-level for meaningful assessment, while the sub-requirements were too granular to provide clarity. The 4 principles and the 7 requirements were therefore favored for the mapping.

*First,* issues mapped to the same key requirement of the EU framework were grouped together to identify and combine related issues from similar groups.

*Then*, the consolidated lists of WG issues for each of the seven requirements were reviewed so commonalities and differences could be identified and discussed before final consolidation.

This helped us find and combine similar issues mapped to different key requirements, which is possible due to the subjective mapping performed by the groups.

A central problem was how to handle the ambiguity of the mapping from issue to key requirement.

We observed that the different groups frequently mapped issues to different key requirements which made the first step of our mapping less effective than was planned.

> Example of mapping to multiple key requirements from [52[
>
> WG technical:
> (*Open vocabulary*) **Low system transparency.**
> It can be difficult to establish a link between input image and output severity score. The system is not easily explainable due to its many blocks and complexities.
>
> (*Closed vocabulary*) Mapped to two key requirements
> **Accountability** > *Human Agency and Oversight*, *Technical Robustness and Safety*,
> **Transparency** > *Explainability, Communication*

In the second step, however, we found similar issues identified by different groups and mapped to different key requirements. (i.e. an issue being mapped to more than one requirement).

To us, this showed that while we agreed on the issues, the different backgrounds provided different perspectives on the underlying problem and its implications.

Similar to the previous step, if an issue was found to be mapped to different requirements, we found a consensus within the group which of them were most applicable, while also accepting that different points of view could lead to different meanings.

2.6.3.2 Use of Natural Language Processing for Consolidation

An explicit goal of the consolidation phase is that the final result should be agreed upon by multiple experts. When the number of issues are relatively small (e.g. less than 10) it is feasible to consolidate them manually [45], and have these results evaluated by other experts.

However, in one of the most complex use cases we assessed [52], [46], we had a very large team of experts (over 40) who identified over 50 issues. Due to a large number of participants, the manual approach was not feasible, as it proved too demanding for a single person to be aware of all issues for consolidation. This in turn also made it difficult to get expert consensus, as we had no initial version to discuss, but only a large list of issues.

We first tried to separate the issues into smaller groups according to the trustworthy AI key requirements they were mapped to, but we quickly found that this approach was not working well, as different WGs assigned similar issues to vastly different key requirements.

The main challenge was that the evaluation work was split into different WGs that tended to use different terminology and jargons popular in their sub-fields (legal, medical, AI/ML/CS, social, bioethics, ethics, etc). This led to situations where the different WGs described using free text similar issues using different terminologies or from a different perspective. This made the consolidation task of mappers difficult since identifying overlaps between such issues is quite challenging. Performing the consolidation manually was both cognitively challenging and time consuming.

To support the process for this use case we built a natural language processing (NLP) tool that uses a deep neural network to map the issue descriptions to numerical vectors in a way that the similarity of the resulting vectors is proportional to the similarity of the descriptions' semantic meaning. This in turn allows us to use established data analysis techniques to identify groups of potentially similar issues. The NLP solution helped us here, as it provided a first separation of issues. This grouping drastically reduced the cognitive effort required to understand the different topics considered in the issues and thereby helped the experts to get a broad view of the issues discussed in other WGs. In addition, with each of the groups sharing semantic topics, it was easy to identify issues that did not belong, and also identify groups that might belong together. This in turn enabled more experts to efficiently participate in the discussion, which in turn led to an identification of the most pressing issues which could incorporate many different viewpoints.

For the assessments performed so far, we have not prioritized the time and resources needed to perform the use case. However, for future assessment we plan to develop a more agile process to reduce the time for an assessment to three months at a maximum.

For that, we plan to experiment with the use of natural language processing for consolidation in future assessments to validate and speed up the approach. However, the tool only enables efficient participation of the mappers and it does not replace the essential human component of the assessment process.

A more extensive explanation of the technical details is available in [47].

## 2.7 Trade-off and Recommendations

The resolve phase completes the process by addressing ethical tensions and by giving recommendations to the key stakeholders. The recommendations might for example concern appropriate use, remedies for mitigating risks, and ability to redress.

One way in which trade-offs and recommendations have been developed in practice is through discussion in the working groups. To give an example, in one of the ethics working groups we listed a set of concrete recommendations and our reasons for highlighting them. For instance when considering the development of an AI system during a pandemic, we must consider how to trade-off standard procedures for securing informed consent against the need for speedy training of an algorithm. We found that it was recommendable that a policy was in place making sure to protect patient rights during a pandemic, where very high societal costs are at stake, and such rights can come under pressure. This recommendation grew out of a more general discussion about ethical issues relating to the use case.

One of the main challenges here is how to motivate and engage with the main stakeholders to make sure that they act upon (some of) the given recommendations. This is an open area of practical research.

## 2.8 Monitor AI over time

It is crucial to monitor that the AI system that fulfilled the trustworthy AI requirement after its initial deployment continues to do so over time. The AI system has multiple factors that evolve over time, the learned model changes its behavior with updated training data or new data for inference, and the software and hardware of the execution environment change, e.g., operating system or Machine Learning library updates. A system once considered trustworthy cannot be guaranteed to remain trustworthy for its lifetime given one of these changes. It is also possible that the "ground truth" or "gold standard" changes if the knowledge in that field changes. Therefore, when required, the resolve phase includes conducting a trustworthy monitoring over time of the AI system (we call it "ethical maintenance") [48]. An initial benefit is constantly updated documentation about the deployed system reflecting its current and past states. As also discussed in the recent proposal for Reward Reports [31], the documentation is beneficial during system maintenance and in the decommissioning phase when the system needs to be replaced or shut down.

## 3. AI Certification and Fundamental Rights Assessment

The certification of AI-based products and services is a growing need for companies wanting to sell products in safety-critical areas [49]. A related requirement for assessment will likely soon be required under the forthcoming AI Act. Our process can assist in the certification process by providing a trustworthy AI assessment of company claims for the system. However, the Z-Inspection® network is not a certifying body under any jurisdiction, nor is the process complete in terms of certification. A fundamental rights assessment for AI systems involving people [50] has recently been proposed as a law by the Dutch parliament. Our process nicely complements such an impact assessment, which is basically a checklist, while our process is a dynamic counterpart which helps verifying claims and identifying ethical dilemmas and tensions from different interdisciplinary perspectives.

## 4. Shortcomings

We recognize that our self-assessments have some limitations, namely:

Both the mappings and the consolidation of the mappings involve subjective decision-making components. ™On the one hand, this clearly shapes the process, as it provides a framework for assessment. On the other hand, this may bear the risk of disregarding other ethical concepts and principles relevant in the context of a use case.

Overall, the assessment is shaped and also limited by the team members' focus on work and expertise. While a very large interdisciplinary team may work on a use case, it is not possible to ensure that every perspective is covered or is equally covered. Z-Inspection®  is a framework that prioritizes inclusion, but it is necessarily limited by the length of time in which the inspection can occur and the skillsets of those interdisciplinary experts who take part.

It is also  important to mention the pros and cons of having experts volunteering time. On the one hand there is no need to pay experts or a consultancy firm (possibly with its own agenda) doing the assessment- but this also means that some experts might have less time to invest and there can be uneven contributions both in quality and quantity. This requires a balanced act to select the experts who are motivated, knowledgeable and have quality time.

## 5. Conclusions

As suggested to us by our Advisory Board member Wonki Min**,** we will consider in our future work to support the following points:

*"From the point of a regulator, there is a need for a framework that enables auditing innumerable AI systems.  Since the audits must be conducted repeatedly over <u>AI systems' lifecycle</u>, how to empower AI regulators to address this issue is the key to secure the effective implementation of AI regulations. In particular, it is critical to design the smart governance structure of auditing that can address institutional and budgetary issues in the process of audit.*

*As we are witnessing the development of AI-as-a-Service, it is also important to build the framework to minimize the accountability dispersion through the technically feasible and socially acceptable allocation of <u>auditing responsibilities among the institutions in the AI value chain.</u> "*

The Z-Inspection® process will necessarily evolve alongside the regulatory environment. While AI remains largely unregulated, Z-Inspection® can provide important validation and ethical consideration in line with guidelines. Once regulation is in place, Z-Inspection® will be able to utilize its experience in assisting AI systems developers and users in navigating the legal and ethical requirements. **Broad and interdisciplinary subject matter expertise will be critical to making a valuable assessment of trustworthiness**, something Z-Inspection® will be able to offer, not only in a self-assessment scenario but, in supplement to third party assessment bodies whose remit will be more focused.

## Acknowledgement

We would like to thank Emma Ruttkamp-Bloem, Mikael Boesen, Helga Brogger and Wonki Min for providing us valuable feedback.

Co-author DV received funding from the European Union's Horizon 2020 research and innovation program under grant agreement no. 101016233 (PERISCOPE), and from the European Union's Connecting Europe Facility program under grant agreement no. INEA/CEF/ICT/A2020/2276680 (xAIM). The funders had no role in study design, data collection and analysis, decision to publish, or preparation of the manuscript.

# Appendix

Over the course of our assessments we found that we were frequently re-using materials from cases as templates. Especially in the early phases, this helped us enormously to get a more structured process and with growing experience, we could further refine the templates to make future assessments even more structured. In the following, we will present a collection of the most useful materials we collected. The templates and code are available at [https://github.com/dennisrv/z-inspection-toolkit](https://github.com/dennisrv/z-inspection-toolkit).

The first useful template is for structuring the kick-off meeting where the team meets the stakeholders for the first time and presents their system. It contains sections for 10 important areas where information is needed, such as the aim of the system, how users interact with the AI system and information on the legal framework and intellectual property. We found that also sending out this template to the owners in advance helped them to structure their presentation of the system and collect the relevant information.

Another useful part for processing the kick-off meeting and evaluating the proceedings was the linking of sections of the recording video to the google docs. Therefore the repository also contains a script to automate this process, along with instructions on how to install and use it.

The last tool contained is a python notebook that illustrates the AI-based approach for the consolidation phase. We used a deep neural network to compute sentence embeddings, mappings of sentences to a high-dimensional numerical vector. This embedding is computed in a way that sentences with similar meaning have a similar embedding. With this we could apply clustering algorithms to identify groups of similar issues, which were then used as a starting point for our consolidation phase. The corresponding publication [47] contains the technical details, as well as a link to code for easy reproduction and additional evaluation of the quality of the resulting groupings.

A question we were asked during the assessment was if the tool works with different languages and if there is a potential bias for non english speaking persons compared to native english speaking persons that the AI NLP does not capture?

At present the tool works for the English language, as this is the language with the most research (and datasets) available. We did not investigate biases regarding non-native English speakers, but in our evaluation the cases where the tool was not working well were not due to subpar use of the english language.

To some extent the model should be robust against non-native speakers, as it is trained in part on English texts crawled from the internet (i.e. forum comments, wikipedia articles, research papers, ...), which are likely to include texts written by non-native speakers.

There is also some effort into so-called "low resource languages", where much less datasets are available (think German, Dutch, Arabic, ...), but we did not try how having everyone writing in their native language would influence the AI. Often there is also a difference between native language and working language.

An interesting question which we would need to address in the future is if there is some tension in having a process for assessing AI which uses another AI which is not assessed.